\documentclass[showpacs,pre,twocolumn,floatfix]{revtex4}
\usepackage{amsmath,amsfonts,amssymb}
\usepackage{graphicx} 
\usepackage{color} 

\begin{document}
\title{Angular velocity distribution of a  granular planar rotator
in a thermalized bath}
\author{J. Piasecki$^{1}$}
\author{J. Talbot$^2$} 
\author{P. Viot$^3$}

\affiliation{$^1$ Institute of Theoretical Physics, University of Warsaw, Ho\.za
 69, 00 681 Warsaw, Poland}

\affiliation{$^2$Department   of  Chemistry  and  Biochemistry,
 Duquesne University, Pittsburgh, PA 15282-1530}
 \affiliation{$^3$Laboratoire de Physique
Th\'eorique de la Mati\`ere Condens\'ee, Universit\'e Pierre et Marie Curie, 4, place
Jussieu, 75252 Paris Cedex 05, France}
\begin{abstract}
The  kinetics   of a  granular   planar  rotator with  a  fixed center
undergoing inelastic collisions with  bath particles is  analyzed both
numerically and analytically by means  of the Boltzmann equation.  The
angular  velocity distribution   evolves  from quasi-gaussian in   the
Brownian limit  to an algebraic  decay in  the  limit of an infinitely
light  particle.   In addition, we   compare this model with  a planar
rotator with a free center.  We propose experimental tests that  might
confirm the predicted behaviors.
\end{abstract}
\maketitle
\section{Introduction}
When macroscopic   particles undergo  inelastic collisions  the  total
kinetic energy decreases with  time. If an  external source of energy,
such  as a vibrating bottom wall,  is present, the  system may reach a
stationary  state.   Despite  similarities  with  equilibrium systems,
however,  equilibrium    statistical mechanical  concepts   cannot  be
applied\cite{G05,BP004,Z06}.   For instance, there is no equipartition
between             different      species   in           polydisperse
systems\cite{WP02,FM02,MP99,HAZ99,BT02},   and velocity  distributions
are, in general, non-gaussian\cite{BCR99,RM00,A005}.

In dilute  systems,  most collisions  only  involve two particles, and
consequently,  a  theoretical description   of the  dynamics has  been
proposed    starting      from     the    Boltzmann       or    Enskog
equation\cite{PB03,BP004}.

The inelastic hard sphere model is a paradigm for granular gases in
the same way that the elastic hard sphere model is for equilibrium
fluids\cite{BP004}. While both account for the exclusion effect,
energy is lost at each collision in the former.  For dissipative
systems, recent studies have shown the relevance of the granular
temperature, even if the absence of equipartition yields a number of
temperatures increasing with the polydispersity or the number of
degrees of freedom of each particle.

Few studies have addressed the effect of particle shape on the
properties of granular gases. In three dimensions, Aspelmeier et
al\cite{HAZ99} showed that the rotational and translational
temperatures are different in the free cooling state of inelastic hard
needles.  Anisotropic tracers (needle and spherocylinder) in a
thermalized bath also display non-equipartition between different
degrees of freedom\cite{VT04,GTV05}.  The purpose of this paper is to
build a simple model which captures the specific features due to the
shape of the particle and the presence of an irreversible microscopic
dynamics.

We present here an investigation of the stationary kinetics of a
planar rotator with a fixed center that undergoes inelastic collisions
with the bath particles.  With these assumptions, the kinetics is
described by a linear Boltzmann equation.  We show that, when the
tracer is much heavier than the bath particle, 
the angular velocity distribution function is quasi-gaussian (the Brownian
limit) whereas it exhibits an algebraic decay in the opposite limit of an
infinitely light granular particle\cite{PV06}.  For all intermediate
cases, there is no simple scaling regime and deviations from gaussian
behavior are captured by analyzing the fourth moment of the
distribution function.  

The      paper is organized  as    follows:  the model, its mechanical
properties  and the  Boltzmann  equation   are presented in    section
\ref{sec:planar-rotator}.  The asymptotic  solution  of the  Boltzmann
equation    in     the    Brownian     limit  is        presented   in
section~\ref{sec:brownian-limit}.  In section
\ref{sec:zero-mass-limit}, analytical results are derived for a
zero-mass  limit and  for  different coefficients  of restitution  and
intermediate cases are considered in section
\ref{sec:intermediate-cases}. In section \ref{sec:models-with-free}
we compare the fixed rotator with one whose center is free, both within
the gaussian approximation. Finally, the 
conclusion discusses possible experimental tests of
our theoretical results. 

\section{The planar rotator}\label{sec:planar-rotator}
\subsection{Definition and mechanical properties}
The model consists of a two-dimensional, infinitely thin needle of
mass $M$, length $L$ and moment of inertia $I=ML^{2}/12$ immersed in a
bath of point particles, each of mass $m$. The needle has a fixed
center of mass, but can rotate freely around its center. It undergoes
instantaneous and inelastic collisions with the surrounding bath
particles.  The motion of the planar rotator can be described with the
angle between a unit vector ${\bf u}$ collinear to the axis of the
needle and the x-axis.

The   rate  of change   of the  orientation
$\dot{{\bf u}}=\omega {\bf u}_\perp$ is equal to  the  angular velocity $\omega \in ]-\infty,+\infty
[$ times a unit vector ${\bf u}_\perp$ perpendicular to ${\bf u}$.

\begin{figure}[t]
\resizebox{8cm}{!}{\includegraphics{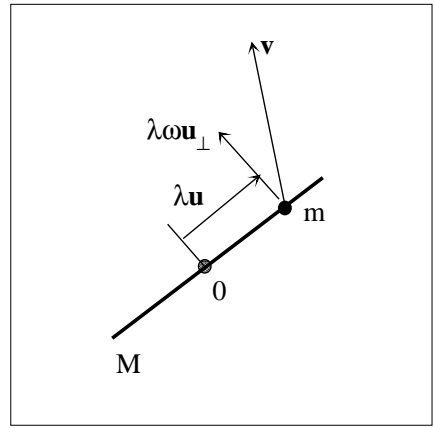}}

\caption{Illustration of a collision between the planar rotator and  a bath particle: 
${\bf u}$ and ${\bf u}^\perp$ are  unit vectors parallel and perpendicular
to the axis of  the rotator.   The
element of the  needle  at  $\lambda {\bf  u}$  moves with  linear  velocity
${\lambda\omega\bf  u}^\perp$.}
\label{fig:1}
\end{figure}

The  angular velocity of the rotator  changes at each binary collision
with a bath particle. The position of the point of impact  along the needle axis is
denoted by $\lambda       {\bf   u}$.  Obviously,    
a condition for collision is
$|\lambda    |<L/2$ (see
Fig.\ref{fig:1}).  The  relative velocity  ${\bf V}$  at the  point of
impact equals
\begin{equation}\label{eq:1}
{\bf V}={\bf v}-\lambda \dot{\bf u} = {\bf v}-\lambda \omega {\bf u}_\perp
\end{equation}
where ${\bf v}$ denotes the velocity of the bath particle.

Due to the dissipative nature of the collision, the relative velocity  changes
according to the collision law
\begin{align}\label{eq:2}
V^*_\perp &=-\alpha V_\perp \\
V^*_{||} &= V_{||} \label{eq:3}
\end{align}
where $0\leq\alpha \leq 1$ is the  normal  restitution coefficient and where  the
indices $\perp$    and   $||$ indicate the  perpendicular     and parallel
components  of any vector relative  to the needle axis, respectively.
When $\alpha=1$,  one recovers an elastic  collision rule.  For the sake of
simplicity,  the tangential component   of  the velocity is  unchanged
during the collision.

Since each collision conserves the total  angular
momentum we have that
\begin{equation}\label{eq:4}
I\omega^* + \lambda m v^*_\perp = I\omega + \lambda m v_\perp ,
\end{equation}
where post-collisional  quantities  are denote with a star.  

By  combining  Eqs.~(\ref{eq:1})-(\ref{eq:3}), the post-collisional bath
 particle velocity is given by 

\begin{equation}\label{eq:5}
v^*_\perp= v_\perp  -\frac{I(1+\alpha)V_\perp }
{I+ m\lambda^2}.
\end{equation} 
whereas the corresponding  post-collisional angular velocity is 
\begin{equation} \label{eq:6}
\omega^*=\omega +\frac{(1+\alpha)V_\perp m\lambda }{I+m\lambda^2}.
\end{equation}
The inverse   transformation  (giving  the pre-collisional  quantities
denoted by a double star) is obtained by substituting $\alpha$ by $\alpha^{-1}$
and the starred quantities by  double-starred quantities.

\subsection{Homogeneous Boltzmann equation}
At low density, one assumes that the needle influences weakly the
local density of the bath and, consequently, that the system remains
homogeneous. After a transient time (not considered here) the kinetics
of the needle becomes stationary and can be described by the
stationary Boltzmann equation. This expresses the invariance  
of the rotator angular velocity distribution
function, $F(\omega)$, resulting from a balance
between collisional gain and loss terms:
\begin{align}\label{eq:7}
 \int_{-L/2}^{L/2} \!\!d\lambda &\int\!    d{\bf v} |v_\perp-\lambda \omega|
 \left(\frac{F(\omega^{**})\Phi_B({\bf v}^{**})}{\alpha^2}-F(\omega)
\Phi_B({\bf v})\right)\nonumber\\&=0,s
\end{align} 
where the pre-collisionnal velocities ${\bf  v}^{**}$ and $\omega^{**}$ are
given   by the   right   hand sides  of   equations   (\ref{eq:5}) and
(\ref{eq:6}), respectively,  with $\alpha$  replaced by $\alpha^{-1}$.
$\Phi_B({\bf v})$ is the time independent bath velocity distribution.
 
The  integration over the   parallel  velocity component $v_{||}$   in
Eq.(\ref{eq:7})  can be readily carried out  since $F$ does not depend
of  this variable.   By changing of  the  integration  variable we can
rewrite Eq.(\ref{eq:7}) as
 
\begin{align}\label{eq:8}
&\int_{-L/2}^{L/2} d \lambda\int dv_\perp |v_\perp| \left[F\left(w+v_\perp\frac{(1+\alpha)m\lambda}{I+m\lambda^2}\right)\right.\nonumber\\&
\left.\phi_B\left(\lambda\omega +v_\perp\frac{(\alpha m\lambda^2-I)}{I+m\lambda^2}\right)
- F(\omega)\phi_B(v_\perp+\lambda\omega )\right]=0,
\end{align}
with $\phi_B(v)= \int dv_{||}\Phi_B(|{\bf v}|)$.

\section{Brownian limit}\label{sec:brownian-limit}
An exact solution     of Eq.(\ref{eq:8})   cannot be obtained       in
general. When the  mass of the planar rotator  is much larger than the
mass of the  bath particle,  however,  one expects that the  deviation
from Maxwellian behavior is weak. It turns out (see Appendix
\ref{sec:Dimensionless}) that exploring the regime corresponding to Brownian motion is equivalent to the analysis of the small $\lambda$
expansion   of the  collision term  (8).   We thus  perform perform  a
perturbative expansion of the integrand of Eq.(\ref{eq:8}) in terms of
$\lambda$. We denote

\begin{align}
G(v_\perp,\omega,\lambda)= \left[F\left(w+v_\perp\frac{(1+\alpha)m\lambda}{I+m\lambda^2}\right)\right.\nonumber\\
\left.\phi_B\left(\lambda\omega +v_\perp\frac{(\alpha m\lambda^2-I)}{I+m\lambda^2}\right)
- F(\omega)\phi_B(v_\perp+\lambda\omega )\right]
\end{align}

with the property that $G(v_\perp,\omega,0)=0$. To  go further, we assume in
the rest of  this section that the bath  distribution,  $\phi_B(v)$, is a
Maxwellian:
\begin{equation}
\phi_B(v)=\phi_M(v)=\sqrt{\frac{m}{2\pi T}} \exp(-mv^2/2T)
\end{equation}

The first  derivative  of $G(v_\perp,\omega,\lambda)$  with  respect to $\lambda$ at  $\lambda=0$
gives the differential equation
\begin{equation}\label{eq:9}
(1+\alpha)\frac{d F(\omega  )}{d\omega }+2 F(\omega ) \frac{\omega I }{T}=0
\end{equation}
whose solution is 
\begin{equation}\label{eq:10}
F(\omega )\propto   \exp\left(-\frac{I\omega^2}{(1+\alpha)T}\right)
\end{equation}

By taking the second derivative, one obtains the differential equation
\begin{align}\label{eq:11}
(1+\alpha)\frac{d^2 F(\omega  )}{d\omega^2 }+\frac{2I}{T}\omega  \frac{d F(\omega  )}{d\omega }+\frac{2I}{T}F(\omega )=0
\end{align}
whose solution is also given by Eq.(\ref{eq:10}).

The third-order expansion gives a polynomial in $v_\perp$ 
\begin{align}\label{eq:12}
\frac{m^3v_\perp^3}{I^3}H_3\left(F(\omega)\right)
-\frac{3m^2v_\perp}{I^2} H_1\left(F(\omega)\right)=0
\end{align}
where
\begin{align}
H_3\left(F(\omega)\right)&=(1+\alpha)^3\frac{d^3 F(\omega  )}{d\omega^3 }+ 3(1+\alpha)^2\frac {I\omega}{T}\frac{d^2 F(\omega  )}{d\omega^2 }+
\nonumber\\
&+3\frac{I}{T}\left((1+\alpha)\frac{I\omega^2}{T}+2\right)\frac{d F(\omega  )}{d\omega }+\nonumber\\
&+\frac{2I^2\omega}{T^2} \left(3(1+\alpha)+\frac{I\omega^2}{T}\right)F(\omega)
\end{align}
and 
\begin{align}\label{eq:13}
H_1\left(F(\omega)\right)&=(1+\alpha)(2+\frac{I\omega^2}{T})\frac{d F(\omega  )}{d\omega }+
\nonumber\\
&+\frac{I\omega}{T} \left((1+\alpha)+2\frac{I\omega^2}{T}\right)F(\omega)
\end{align}

The solution of $H_3=0$ is given by Eq.(\ref{eq:10}) again, but the solution of $H_1=0$ is  
\begin{equation}\label{eq:14}
F(\omega )\sim  \left(2+\frac{I\omega^2}{T}\right)^{\frac{1-\alpha}{1+\alpha}}\exp\left(\frac{-I\omega^2}{(1+\alpha)T}\right)
\end{equation}
 
If  Eq.(\ref{eq:14})   if  different  from  Eq.(\ref{eq:10}),    it is
interesting to note that this solution is  the same Maxwellian times a
slow decreasing function  (If $\alpha=1$, Eq.(\ref{eq:14}) is identical  to
Eq.(\ref{eq:10})).  The fourth  derivative  of  $G(v_\perp,\omega,\lambda)$ has  been
calculated  with    the  software Maple, and     it   occurs that  the
differential    equation associated  with     the $v_\perp^3$ term  has  a
Maxwellian solution  but the  solution  of the  differential equations
associated with the $v_\perp$-term  is  given by the  Maxwellian  solution
multiplied by a slow varying  function.  Therefore, we conjecture that
the  complete  solution  is    given   by Eq.(\ref{eq:10})  times    a
sub-dominant term.

\begin{figure}[t]

\resizebox{8cm}{!}{\includegraphics{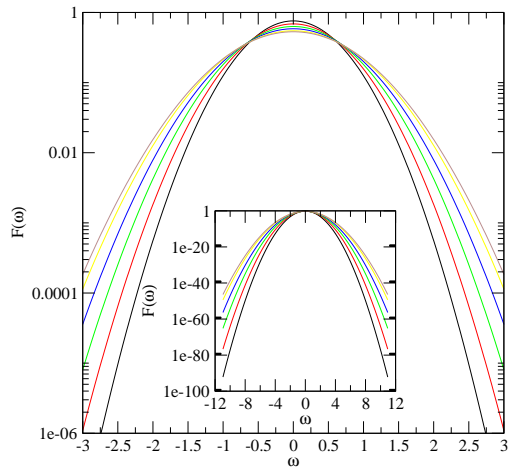}}\\

\resizebox{8cm}{!}{\includegraphics{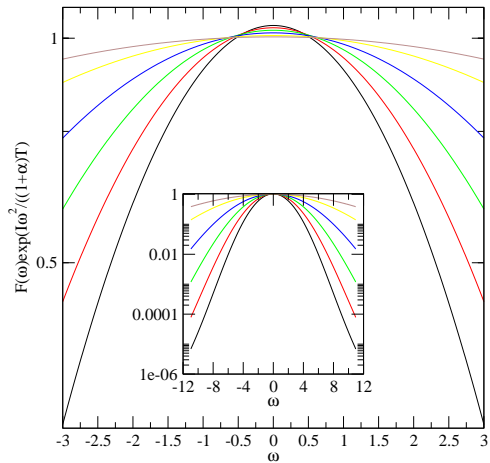}}
\caption{(a) Log-linear plot $F(\omega)$    for $M/m=10$ and with different values
 of the coefficient of restitution $\alpha=0,0.2,0.4,0.6,0.8,0.9$ (from top
 to bottom, center).(b) Log-linear  plot $F(\omega)\exp(I\omega^2/((1+\alpha)T))$ for $M/m=10$. The insets display the same curves on a larger range of angular velocities.
 }\label{fig:2}
\end{figure}

In  order to  check this assumption,   we have solved  numerically the
Boltzmann  equation.   Since   Eq.(\ref{eq:8})   is linear   and   the
distribution $F(\omega)$  is  a  one-variable  function, we have   used an
iterative method  that is very efficient and provides  a much more accurate
solution\cite{BMP02} than a DSMC method\cite{MS00}.

By   using Eq.(\ref{eq:8}),   the  numerical resolution    is obtained by
iterating the following equation
\begin{align}
F^{(n+1)}(\omega)=&C(\omega)\int          d    \lambda  \int        dv
|v|F^{(n)}\left(w+v\frac{(1+\alpha)m\lambda}{I+m\lambda^2}\right)\nonumber\\
&\phi_M\left(\lambda\omega                             +v\frac{(\alpha
m\lambda^2-I)}{I+m\lambda^2}\right)
\end{align}
where 
\begin{equation}
C(\omega)=\left(\int d \lambda \int dv |v-\lambda\omega |\phi_M(v)\right)^{-1}
\end{equation}
$C(\omega)$ is an explicit function when the bath particle
distribution $\phi_M(v)$ is a Maxwellian.  The velocity distribution
is sampled on a one-dimensional grid with $1000$ points. The
integrations over $\lambda$ and $\omega$ are performed with
Simpson's rule with $100$ and $1500$ points, 
respectively.  For values of the velocities which do not match to the
lattice points a linear interpolation is performed.  The initial
distribution is taken as the Maxwellian, Eq.(\ref{eq:10}). Except when
the mass of the planar rotator is extremely small, the method
converges very rapidly.

\begin{figure}[t]

\resizebox{8cm}{!}{\includegraphics{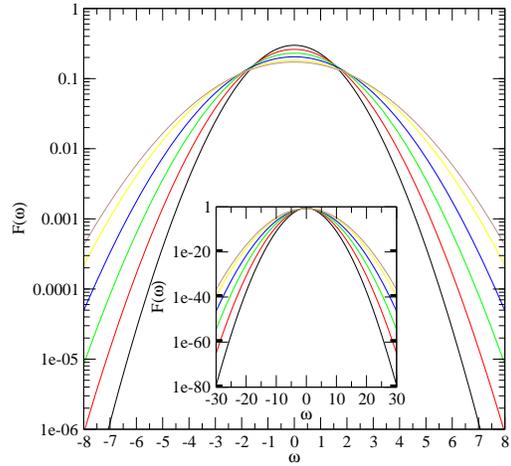}}\\

\resizebox{8cm}{!}{\includegraphics{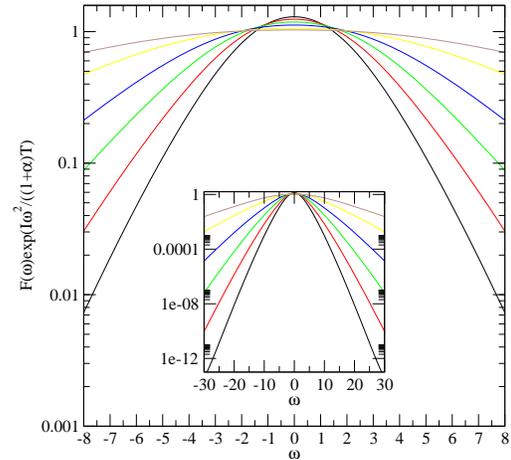}}
\caption{Same caption as Fig.2 for   for $M/m=1$ 
 }\label{fig:3}
\end{figure}

Fig.\ref{fig:2}a displays the angular velocity distribution
$F(\omega)$ for a mass ratio $M/m=10$ for different values of the
coefficient of restitution $\alpha$.  The deviations from the
Maxwellian,  shown in Fig.\ref{fig:2}b, increase with decreasing
coefficient of restitution, but compared to the function
$F(\omega)$, they vary weakly with the angular velocity.
The insets show that the iterative method allows the tails of the
distribution function to be obtained with high precision (unlike with 
the DSMC method).
Fig.\ref{fig:3} shows the angular velocity distributions for a 
mass ratio equal to one. The 
deviations from the gaussian behavior are more pronounced than those
for a mass ratio of $10$, but they are still negligible 
compared to the the leading gaussian term.

\section{Zero-mass limit}\label{sec:zero-mass-limit}
When both  the mass of the  needle and the coefficient  of restitution
are  equal to zero,   the Boltzmann equation, Eq.(\ref{eq:8}), can  be
solved exactly.  Let us  review  the  special characteristics of  this
limiting case. First, because of the zero mass of the tracer particle,
the velocity of  the bath particles never changes  as the result  of a
collisions.In addition, the   angular velocity of  the needle  has  no
memory  of its pre-collisional value.    This quantity is reset  after
each   collision,  acquiring instantaneously  the  value  $v_\perp /\lambda$. By
introducing   the  variable $v_\perp =  (\lambda\omega)    y$, Eq.(\ref{eq:8}) can be
written as

\begin{align}\label{eq:15}
&\int_{-L/2}^{L/2} d \lambda \lambda^2\int dy |y|F\left(w\left(1+y\frac{(1+\alpha)m\lambda^2}{I+m\lambda^2}\right)\right)\nonumber\\&
\phi_B\left(\lambda\omega \left(1+y\frac{(\alpha m\lambda^2-I)}{I+m\lambda^2}\right)\right)\nonumber\\&
= F(\omega)\int_{-L/2}^{L/2} d \lambda \lambda^2\int dy |y-1 |\phi_B(\lambda\omega y )
\end{align}

When $I=0$ and $\alpha=0$, this simplifies to

\begin{figure}[t]

\resizebox{8cm}{!}{\includegraphics{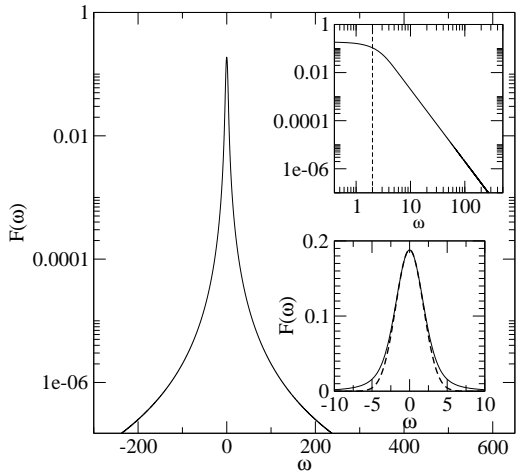}}
\caption{Log-linear plot $F(\omega)$ of the ``zero mass''  
needle with $\omega_c=2$. The  upper inset  displays the log-log  plot  showing  the power-law
decay for sufficiently large $\omega $ values (the vertical dashed line corresponds to $\omega=\omega_c$). The lower inset is a log-log plot showing the gaussian behavior for $\omega<\omega_c$ (the dashed curve corresponds to the gaussian approximation)}\label{fig:4}
\end{figure}

\begin{align}\label{eq:16}
&\int_{-L/2}^{L/2} d \lambda \lambda^2\int dy |y|F\left(w\left(1+y\right)\right)
\phi_B\left(\lambda\omega\right) \nonumber\\&
= F(\omega)\int_{-L/2}^{L/2} d \lambda \lambda^2\int dy |y+1 |\phi_B(\lambda\omega y )
\end{align}
The exact solution $F(\omega )$ of Eq.(\ref{eq:16}) is obtained by integrating
the bath distribution $\phi_B(\lambda \omega )$  weighted by the 
position $\lambda$ of the  point of impact:
\begin{equation}\label{eq:17}
 F(\omega)=\int_{-L/2}^{L/2} d\lambda \left(\frac{2\lambda}{L}\right)^{2}\phi_B(\lambda \omega )
\end{equation}

Note that this solution is independent of specific assumptions about
the distribution of the bath particles.  If we make the weak
assumption that the second moment of the distribution $\phi_B(v)$ is
finite, i.e. that the bath is characterized by a finite granular
temperature, it can be shown that $F(\omega)$ decays algebraically as
$\omega^{-3}$.  The long tail of $F(\omega)$    arises from collisions
near the center of the rotator that result in large angular
velocities.  It is worth noting that solutions of the Bolzmann
equation with power-law decay exist  for isotropic particles, but not
with a thermalized bath of particles\cite{BM05,BMM05}.

While the granular  temperature of the  bath particles is finite, that of
the tracer is not well defined since its zero mass
implies an infinite mean squared angular  velocity.  This difficulty is
removed when the granular particle has a small, but finite mass (see
below)

An explicit expression of the angular velocity distribution function
$F(\omega)$ can be obtained when the
bath particles have a Maxwellian velocity distribution,
$\phi_B(v)=\sqrt{m/2\pi T}\exp(-mv^2/2T)$. In this case

\begin{align}\label{eq:18}
F(\omega)&=-\sqrt{\frac{1}{\pi}}\frac{\omega_c}{\omega^2}\exp\left(-\frac{\omega^2}{\omega_c^2}\right)+
\frac{\omega_c^2}{2\omega^3} {\rm erf}\left(\frac{\omega}{\omega_c} \right)
\end{align}

where $\omega_c=\sqrt{8T/(mL^2)}$  is a crossover frequency between
two regimes. For  $\omega >\omega_c$, $F(\omega)$ has  a power-law behavior and for $\omega
<\omega_c$, $F(\omega)\sim \exp(-3\omega^2/(5\omega_c^2))$ is gaussian (see Fig.\ref{fig:4}).
It  is important to  note that  when  the power-law regime begins, the
amplitude  of  $F(\omega)$  has only decreased  by   an order of  magnitude
compared to the maximum, $F(0)$.

An obvious question is whether the behavior just described persists
when either $\alpha$ or $M/m$ is different from zero.  Is the power law regime
sustained in these cases?

We first consider the case where the mass ratio is maintained at zero,
but the coefficient of restitution is allowed to take all values
between $0$ and $1$.  For $\alpha=1$ (elastic collisions) the solution
of the Boltzmann equation is the expected Maxwell distribution:
\begin{equation}\label{eq:19}
F(\omega)= \sqrt{\frac{I}{2\pi T}}\exp\left(-\frac{I\omega^2}{2 T}\right)
\end{equation}
but the limit $I\to 0$ does not  yield a probability distribution.

\begin{figure}[t]
\resizebox{8cm}{!}{\includegraphics{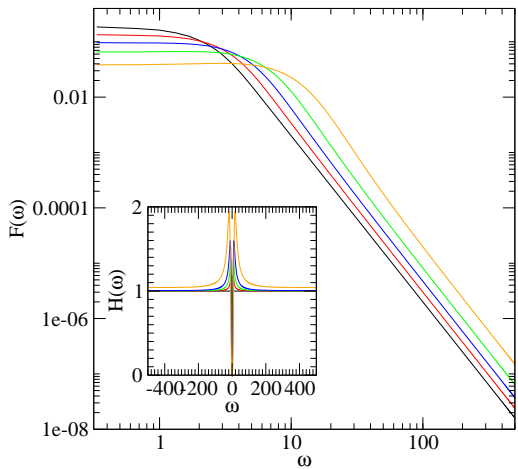}}
\caption{Log-Log plot of $F(\omega)$  of the ``zero mass''  needle with different values 
of the coefficient  of  restitution $\alpha=0,0.2,0.4,0.4,0.8$ from  top to
bottom.  The inset shows  the $\omega$-dependence of the  rescaled function
$H(\omega)=F(\omega)\frac{mL^2(1-\alpha)}{4T(1+\alpha)} \omega^{3}$  for $\omega_c=2$.}\label{fig:5}
\end{figure}

For $0<\alpha<1$ and $ I=0$, we were unable to obtain an analytical
solution and we investigated the behavior numerically with the method
described above.  Figure \ref{fig:5} shows the logarithm of the
distribution function versus the angular velocity.  As in the case
$\alpha=0$, two distinct regimes characterize $F(\omega)$: a scaling
regime for $\omega>\omega_c(\alpha)$, where $\omega_c(\alpha)$ is a
cutoff which increases with $\alpha$ and a gaussian behavior at low
frequencies.

Fig.\ref{fig:5} shows that, in the power-law regime, while the
amplitude of $F(\omega)$ decreases when $\alpha$ increases, 
the exponent of the power-law is independent of
$\alpha$.  It is possible to obtain this result analytically by means
of an asymptotic analysis of the Boltzmann equation, Eq.(\ref{eq:16}).
Details of the calculation are given in
Appendix~\ref{sec:asympt-behav-angul}. The final result, for $M=0$ and
$0 < \alpha < 1$, is
\begin{equation}\label{eq:20}
F(\omega)\sim \frac{4T(1+\alpha)}{mL^2(1-\alpha)\omega^{3}}.
\end{equation}

Unlike the case where $\alpha=0$, the rotator has a  memory of its previous
angular velocity  after  a collision  with a bath  particle,  $\omega^*=\alpha \omega
+(1+\alpha)v_\perp/\lambda$.   This  memory  effect can  be   very small  when a bath
particle collides  near the center  of the rotator (small  $\lambda$), which
explains why, for  large angular velocities, the distribution function
behaves similarly to the case $\alpha=0$.

The            inset      of       Fig.\ref{fig:5}            displays
$H(\omega)=F(\omega)\frac{mL^2(1-\alpha)}{4T(1+\alpha)} \omega^{3}$    versus $\omega$, showing that
the asymptotic  behavior is  rapidly  reached  and that the  numerical
results agree accurately with Eq.(\ref{eq:20}).

\section{Intermediate cases}\label{sec:intermediate-cases}
We have considered  in the preceding sections the two  limiting
cases  of a large mass  of the planar  rotator (Brownian limit), where
the angular velocity  distribution  function displays a  gaussian-like
behavior and a zero  mass case where, surprisingly,  a solution of the
Boltzmann  equation exists with a power-law  decay of the distribution
function. In this section, we investigate the 
intermediate case that corresponds to most physical situations.

We consider  a planar rotator with a small  but finite mass $0<
M \ll m$.  In  Eq.(\ref{eq:15}), since the integrand is an 
even function of $\lambda$, integration can be restricted to the positive 
abscissa. Moreover, the integration can be divided into two parts: 
\begin{equation}
\int_0^{L/2}d\lambda  =\int_0^\epsilon d\lambda  +\int_\epsilon^{L/2}d\lambda 
\end{equation}
where $\epsilon>>\sqrt{\frac{I}{m}}$  (but  $\epsilon$ being  always  small compared
with $L/2$). For small angular velocities $\omega$, the contribution of the
first integral vanishes, and performing a first-order expansion of the
arguments of the integrand, one obtains

\begin{equation}
\int_\epsilon^{L/2}d\lambda F(\omega (1+y(1+\alpha)))\phi_B(\lambda (1+\alpha y))
\end{equation}
which leads to the  Boltzmann equation of  the massless  particle when
$\epsilon \to 0$.

The existence of a finite mass allows  for restoring a finite granular
temperature.  For  small    mass $M$,  as  shown  above,   the angular
distribution function $F(\omega)$ has a power  decay regime truncated at an
upper $\omega_c^{(2)}$ where the  gaussian begins. The granular temperature
given   by the product   of the momentum  of  inertia times the second
moment of the  distribution function is  dominated by the integral  in
the power-law regime.
\begin{align}
T=\int d\omega F(\omega)I\omega^2 \nonumber\\
\simeq \int_{-\omega_c^{(2)}}^{\omega_c^{(2)}}I\frac{d\omega }{\omega}
\sim I\ln(I)
\end{align}

which means that the granular temperature of small-mass planar rotator
goes to zero when the mass decreases (even if the quadratic average of
the angular velocity diverges logarithmically).

\begin{figure}[t]
\resizebox{8cm}{!}{\includegraphics{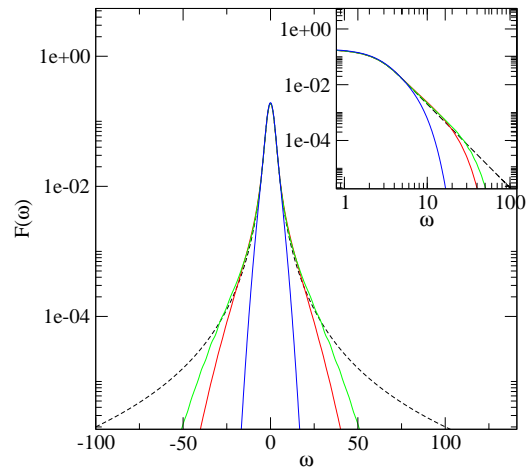}}
\caption{Log-Log plot of $F(\omega)$ for a needle with a mass $M/m=0.005,0.01,0.1$
(dashed curve corresponds to $M=0.0$). 
The   inset   shows  the  crossover  between   the power  law  and the
gaussian-like asymptotic}\label{fig:6}
\end{figure}

Fig.\ref{fig:6} shows the distribution function for three small mass ratios $M/m=0.005,0.01,0.1$ with $\alpha=0$. As  expected the low frequency
distribution  is well   approximated  by   the massless   distribution
function.   The  inset shows that the range of the 
power-law decay 
decreases as the mass of  the  planar rotator  increases. For large
angular   velocities,  the distribution   function resumes the gaussian
behavior, $F(w)\sim\exp(-I\omega^2/((1+\alpha)T))$, irrespective of 
the needle-to-bath particle mass ratio.

In summary, the   velocity distribution keeps  memory  of the massless
solution      up      to  second     crossover    angular     velocity
$\omega_c^{(2)}\sim\sqrt{2T/I}$.     If the mass    of the  planar  rotator is
sufficiently small, one  observes three successive regimes:  first, a
gaussian  decay, second a power-law and  finally a gaussian-like decay
(sub-dominant terms are present).

\begin{figure}

\resizebox{8cm}{!}{\includegraphics{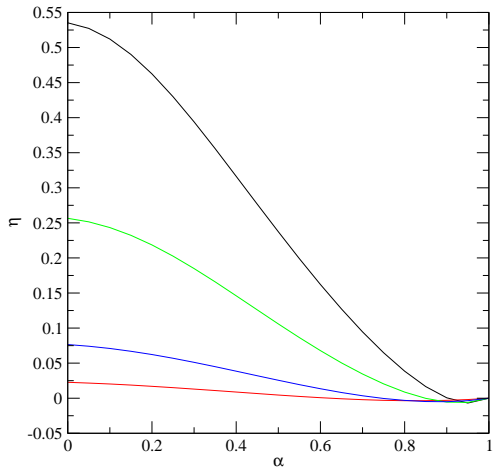}}

\caption{Deviations of the angular velocity distribution function from a gaussian for  $M/m=0.1,1,2,10$, top to bottom.}\label{fig:7}
\end{figure}

When the masses of the planar rotator and a bath particle are
comparable, the two crossover angular velocities merge and the
power-law regime disappears. However, $F(\omega)$ may still deviate
significantly from a gaussian. In order to show this we introduce the
quantity
\begin{equation}
\eta=\frac{\langle \omega^4\rangle }{3\langle \omega^2\rangle^2 }-1
\end{equation}
which is zero for a   gaussian  distribution.    Fig.\ref{fig:7}
displays $\eta$  as a function of the  coefficient of restitution $\alpha$ for
different masses $M/m=0.1,1,2,10$.  

For the spherical tracer particle, the distribution function is a pure
gaussian   when      the         bath  of   particles           is   a
gaussian\cite{MP99}. Deviations from the gaussian occur, however, in a
mixture composed of granular particles of finite density immersed in a
thermostat   providing the  energy   via elastic  collisions when this
energy  is   redistributed   within the  granular    component through
inelastic  encounters\cite{BMP02}.   For  anisotropic   particles, our
calculations show  that  deviations are present  even  in the limiting
case of infinite dilution.

\section{Models with free or fixed center}\label{sec:models-with-free}
The model  of  granular   planar  rotator  with a  free  center   in a
thermalized bath   has    been previously   investigated  by  two   of
us\cite{VT04}. In this model, the  tracer particle has, in addition to
the rotational one, two translational degrees of freedom.

By  using an approximate theory, as  well as  numerical simulations of
the  Boltzmann   equation, it was shown   that   the translational and
rotational  granular   temperatures are both  smaller    than the bath
temperature and  also  different from each  other.    In addition, the
translational   and     rotational      degrees   of   freedom     are
correlated\cite{TV06}.

In this section, we   compare the rotational granular temperatures for
the two models, all parameters  being the same (mass, coefficient of
restitution, bath temperature,...).

In order    to  obtain an analytical    expression  for the rotational
temperature, we use   a method originally  proposed  by  Zippelius and
colleagues  that consists of   calculating  the second moment of   the
angular velocity of  the Boltzmann equation\cite{HAZ99,AHZ00,VT04}. In
a stationary state, this  quantity is constant,  and by using an gaussian
ansatz for $F(\omega)$
\begin{equation}
F(\omega)\propto \exp\left(\frac{-I{\bf \omega}^2}{2\overline {T}}\right),
\end{equation}
one obtains a  closed  equation for the granular temperature
$\overline{T}$ as a function of  microscopic quantities:

\begin{align}\label{eq:21}
\int_0^1dx                        \frac{\overline{T}}{T}x^2\frac{\sqrt{1+
\frac{\overline{T}}{T}kx^2}}{1+kx^2}& =\frac{1+\alpha}{2}\int_0^1dx
\frac{x^2(1+\frac{\overline{T}}{T}kx^2)^{3/2}}{(1+kx^2)^2},
\end{align}
where 
\begin{equation}
k=\frac{mL^2}{4I}
\end{equation}
Details      of    the      calculation         are     given       in
Appendix~\ref{sec:gran-rotat-temp}.  Eq.~(\ref{eq:21}) is an  implicit
equation for $\overline{T}$, but, for a given  value of $\alpha$ and of the
mass ratio, it can be solved with standard numerical methods.

\begin{figure}
\resizebox{8cm}{!}{\includegraphics{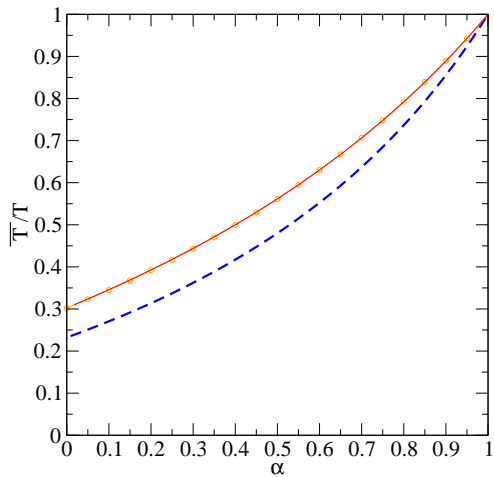}}
\caption{Effective granular temperature $\overline{T}$ as a function of
 the restitution of coefficient $\alpha$ for  a fixed (full curve) and free
 rotator  (dashed  curve)  in   a  bath of  point particles.   Circles
 correspond to  the exact temperatures  by computing the second moment
 of  the distribution function obtained   by the numerical resolution.
 The mass ratio is $M/m=1$ }\label{fig:8}
\end{figure}

Fig~\ref{fig:8} shows the variation of the rotational temperature with
the  normal coefficient of restitution  for a mass  ratio of $1$ for a
fixed  (solid curve) and  free  (dashed  curve) planar rotator.    The
temperature is always  higher when the  center is fixed  except in the
case of  elastic collisions, $\alpha=1$.   The   circles correspond to  the
``exact'' temperatures obtained by computing  the second moment of the
distribution  function $F(\omega)$, which  shows    that the above   method
provides accurate approximate results for estimating  the granular temperatures.

\begin{figure}
\resizebox{8cm}{!}{\includegraphics{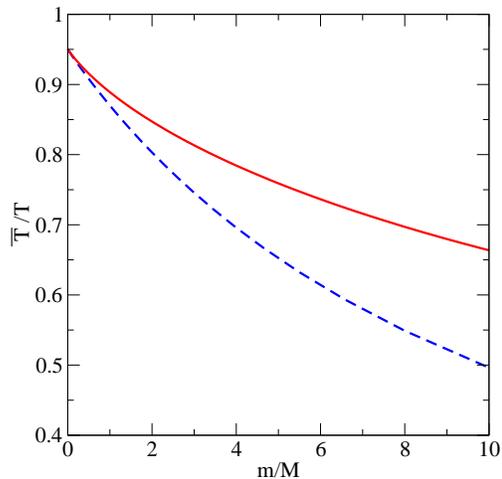}}
\caption{Effective granular temperature $\overline{T}$ as a function of 
the mass ratio  $M/m$ for a fixed (full curve) 
and free rotator (dashed curve) in a bath of point particles. The coefficient of restitution is $\alpha=0.9$.}\label{fig:9}
\end{figure}

We also  consider the variation of  the granular temperature  with the
mass ratio   for  a given value of    the restitution coefficient: See
Fig~\ref{fig:9}.  It   is easy to  verify that  in the Brownian limit,
i.e.   when  the mass  of the rotator  is much  larger of the particle
mass, the granular  temperature goes  to a  value, $(1+\alpha)/2$, that  is
independent of the mass ratio   and, probably, the shape of  particle.
As the  mass   of the needle   decreases, the  difference  between the
rotational  temperatures  of   the free   and   fixed  planar  rotator
increases.  For elastic   systems, the rotational temperatures  remain
identical  for the two different  situations  (free or fixed center of
mass),  but significant  differences  occur for  a  granular particle.

This phenomenon is more pronounced with varying mass ratio than with
varying coefficient of restitution. Therefore, by monitoring the
rotational motion of a granular needle in a bath of significantly heavier
particles in two successive experiments (free and fixed center of mass),
it should be possible to observe the absence of equipartition for a single
particle.

\section{Conclusion}\label{sec:conclusion}

We have shown that  the stationary angular  velocity distribution of a
planar rotator with  a fixed center  that collides inelastically  with
particles in a thermalized bath displays  a variety of behavior as the
mass of the rotator is changed.  Starting from a quasi-gaussian regime
when the rotator is much heavier  than the bath particle, $F(\omega)$ shows
significant deviations from the gaussian  when the mass of the rotator
is  comparable to   that of  a   bath particle. As    the rotator mass
decreases further, an intermediate power-regime also appears.

We believe that these features are not specific to this simple model,
but are generic for all granular systems containing non-spherical
particles.  Furthermore, we expect that the non-gaussian character
that is present in the idealized configuration of a thermalized bath
will be amplified in in the presence of a granular (non-thermal) bath.
Several experiments on intensely vibrated granular systems using high
speed photography\cite{FM02,FM04}, image analysis and particle
tracking\cite{WHP01,WP02} have shown that many quantities, including
granular temperature and velocity profiles, can be precisely measured.
As recent experimental studies have demonstrated, some of the same
techniques can also be applied to granular
rods\cite{BNK03,VKT04,GHSLN06,LV04,LV06}. We believe that the
significant difference between the granular temperatures of a free and
fixed rotator that we have identified (Fig~\ref{fig:9}) should be
detectable using available experimental methods.
\begin{acknowledgments}
J.P.  acknowledges  the  Ministry   of Science  and   Higher Education
(Poland) for financial support  (research project N20207631/0108).  We
thank Alexis Burdeau for helpful discussions.
\end{acknowledgments}

\appendix
\section{Dimensionless Boltzmann equation}\label{sec:Dimensionless}
In order to consider  the Brownian motion regime,  it is convenient to
introduce dimensionless variables $ \Omega =
\sqrt{I/k_{B}T}\omega $ and $ u_{\perp} = \sqrt{m/k_{B}T}v_{\perp} $.

In terms of these variables the integrand in the Boltzmann equation (8)
takes the form

\begin{align}\label{eq:22}
F\left( \Omega + u_{\perp}\frac{(1+\alpha)\epsilon}{1 + \epsilon^{2}}\right)
\phi_{B}(\Omega\epsilon +u_{\perp}\frac{\alpha\epsilon^{2} - 1}{1 + \epsilon^{2}}  )\nonumber\\-F(\Omega)\phi_B(u_\perp +\epsilon \Omega )
\end{align}

where $ \epsilon = \sqrt{m\lambda^{2}/I} = \sqrt{12 m/M}\lambda/L$.

It is thus clear by inspection that the expansion of the dimensionless
collision term (\ref{eq:22}) in powers of $\epsilon $
is equivalent to the expansion of the collision term (8) in powers of
$\lambda$.  Notice that keeping the variable $\Omega$ fixed when exploring
the region of $m \ll M$ corresponds to the Brownian motion asymptotics, as
then the rotational energy  $I\omega^{2}$ is maintained at a fixed ratio
with the thermal energy $k_{B}T$.
\section{Asymptotic behavior of the angular velocity distribution in the zero mass  limit}\label{sec:asympt-behav-angul}
Performing the change of variable $u= \omega( y+1)$ in the
left-hand side of Eq.(\ref{eq:15}) and  $u=\omega y$ in the right-hand side,
one obtains
\begin{align}\label{eq:23}
\int_{-L/2}^{L/2} d \lambda \lambda^2\int& du |u-\omega |F\left((1+\alpha)u-\alpha \omega \right)\nonumber\\
&\phi_B\left(\lambda ((1-\alpha)\omega +\alpha u)\right)\nonumber\\
&= F(\omega)\int_{-L/2}^{L/2} d \lambda \lambda^2\int du |u-\omega  |\phi_B(\lambda u )
\end{align}

When $|\omega|\to\infty $, one has 
\begin{equation}
\int du |u-\omega  |\phi_B(\lambda u )\sim \frac{|\omega |}{\lambda}
\end{equation}
and therefore, the right-hand side of Eq.(\ref{eq:23}) becomes
\begin{equation}
F(\omega)\left(\frac{L}{2}\right)^2|\omega|
\end{equation}
Performing a  similar   analysis for      the  right-hand side      of
Eq.(\ref{eq:23}), one gets  the  asymptotic relation of the  Boltzmann
equation (for the zero-mass limit)
\begin{align}\label{eq:24}
F(\omega)\left(\frac{L}{2}\right)^2=\int_{-L/2}^{L/2} d \lambda \lambda^2\int& du F\left((1+\alpha)u-\alpha \omega \right)\nonumber\\
&\phi_B\left(\lambda ((1-\alpha)\omega +\alpha u)\right)
\end{align}
Let us introduce the Fourier  transforms of the distribution functions
$F(\omega)$   and   $\phi_B(u)$   (with the      convention  $\hat{F}(k)=\int  d\omega
F(\omega)e^{-ikr}$).  The   right-hand-side   of   Eq.(\ref{eq:24})   can  be
expressed as
\begin{align}\label{eq:25}
\int \frac{dk}{2\pi}\int \frac{dq}{2\pi}\int_{-L/2}^{L/2} d \lambda \lambda^2\hat{F}(k)\hat{\phi}_B(q)\nonumber\\
\int du e^{ik[(1+\alpha)u-\alpha \omega]}e^{iq\lambda[(1-\alpha)\omega+\alpha u]}
\end{align}
By using  the property $\int dx  e^{iax}=2\pi\delta(a)$, integration over $v$ in
Eq.(\ref{eq:25}) can be carried out and one obtains

\begin{align}\label{eq:26}
\int \frac{dq}{2\pi}\int_{-L/2}^{L/2} d \lambda \lambda^2\hat{F}\left(\frac{\lambda \alpha q}{1+\alpha }\right)\frac{\hat{\phi}_B(q)}{1+\alpha}e^{i\omega\left(\frac{ \lambda  q}{1+\alpha }\right)}
\end{align}
By  taking the Fourier  transform   of Eq.(\ref{eq:24}) and by   using
Eq.(\ref{eq:26}),  the   asymptotic form   of  the Boltzmann  equation
becomes
\begin{align}\label{eq:27}
\hat{F}(k)\left(\frac{L}{2}\right)^2=\hat{F}(\alpha k)\int_{-L/2}^{L/2} d \lambda |\lambda|\hat{\phi}_B\left(\frac{1+\alpha }{\lambda }q\right)
\end{align}
The integral of the right-hand side of Eq.(\ref{eq:27}) can explicitly
performed
\begin{align}\label{eq:28}
\int_{-L/2}^{L/2}  d \lambda |\lambda|\hat{\phi}_B\left(\frac{1+\alpha       }{\lambda      }q\right)&=\int_0^1  d\mu
\exp\left(-2\frac{(1+\alpha)^2k^2T}{\mu               mL^2}\right)\nonumber\\
&=\exp\left(-2\frac{(1+\alpha)^2k^2T}{
mL^2}\right)\nonumber\\&-2\frac{(1+\alpha)^2k^2T}{
mL^2}E_i\left(1,2\frac{(1+\alpha)^2k^2T}{ mL^2}\right)
\end{align}
where $E_i(1,x)$ is the exponential integral.
For small values of $k$, Eq.(\ref{eq:28}) behaves as
\begin{align}\label{eq:29}
\int_{-L/2}^{L/2} d \lambda \lambda\hat{\phi}_B(\frac{1+\alpha }{\lambda }q)\simeq 1+4\frac{(1+\alpha)^2T}{mL^2}k^2\ln( |k| )
\end{align}
Inserting Eq.(\ref{eq:29}) in Eq.(\ref{eq:27}) yields
\begin{equation}\label{eq:30}
\hat{F}(k)\left(\frac{L}{2}\right)^2=\hat{F}(\alpha k)( 1+4\frac{(1+\alpha)^2T}{mL^2}k^2\ln( | k | )
\end{equation}
Iterating Eq.(\ref{eq:30}) and by using that $\hat{F}(k)=1$, one obtains that
\begin{equation}\label{eq:31}
\hat{F}(k)= 1+\frac{1+\alpha}{1-\alpha}\frac{4T}{mL^2}k^2\ln( | k | )
\end{equation}
The inverse Fourier transform of Eq.(\ref{eq:31}) leads to Eq.(\ref{eq:20})

\section{Granular rotational temperature of the planar rotator with a gaussian approximation}\label{sec:gran-rotat-temp}
By taking the second moment of Eq.(\ref{eq:7}), one obtains the following equation
\begin{align}\label{eq:32}
\int_{-L/2}^{L/2}d \lambda \int v \int d\omega \theta(\omega\lambda -v) |v-\omega \lambda|F(\omega)\phi_B(v)\Delta\omega^2=0
\end{align}
where $\Delta \omega^2=\omega^{*2}-\omega^2$.  This equation means that for a
stationary   state   the second  moment   of  the distribution is time
independent, or in other words, that the loss of the rotational energy
of the planar  rotator induced by  inelastic collisions is compensated
on average by collisions with bath particles with higher velocities.

By using Eq.(\ref{eq:6}),   the difference between the  square angular
velocities at a collision is given by
\begin{align}
\Delta \omega^2&=-\lambda(1+\alpha)\frac{{\bf V}.{\bf u}^\perp(\omega^*+\omega_)}
{\frac{I}{m}+\lambda^2}\nonumber\\
&=-2\lambda(1+\alpha)\frac{{\bf V}.{\bf u}^\perp\omega}
{\frac{I}{m}+\lambda^2}+
\lambda^2(1+\alpha)^2\frac{({\bf V}.{\bf u}^\perp)^2}
{\left(\frac{I}{m}+\lambda^2\right)^2}.
\end{align}
Introducing the dimensionless vectors
\begin{align}
{\bf s}=&(s_x,s_y)\nonumber\\
&=\left(\sqrt{\frac{m}{2T}}v,\sqrt{\frac{I}{2\overline{T}}}\omega\right)
\end{align}
and 
\begin{align}
{\bf G}=&(G_x,G_y)\nonumber\\
&=\left(\sqrt{\frac{2T}{m}},\sqrt{\frac{2\overline{T}}{I} }\lambda\right)
\end{align}
Therefore the scalar product ${\bf G}.{\bf s}$
gives
\begin{equation}
{\bf G}.{\bf s}=v-\omega\lambda 
\end{equation}

Eq.(\ref{eq:32}) can be expressed as
\begin{align}\label{eq:33}
&\int_{-L/2}^{L/2}d \lambda\int d{\bf s}\exp(-{\bf s}^2)\theta(-{\bf G}.{\bf s})|{\bf G}.{\bf s}|\nonumber\\
&\left[-\frac{2\lambda(1+\alpha){\bf G}.{\bf s}}{\frac{I}{m}+\lambda^2}\sqrt{\frac{2\overline{T}}{I}}s_y+
\frac{\lambda^2(1+\alpha)^2({\bf G}.{\bf s})^2}{\left(\frac{I}{m}+\lambda^2\right)^2}\right]=0
\end{align}
Let us define a new coordinate system\cite{AHZ00}  where the y-axis is parralel to
${\bf G}$: units vectors are  denoted $({\bf e}_1,{\bf e}_2)$  whereas
the  unit vectors of  the original system  $({\bf e}_x,{\bf e}_y)$. It
follows that ${\bf G}=|{\bf G}|{\bf e}_1$ and therefore one can write that
\begin{equation}\label{eq:34}
{\bf G}{\bf e}_1{\bf e}_y=|{\bf G}|{\bf e}_x=G_x
\end{equation}
Inserting  Eq.(\ref{eq:34})  in Eq.(\ref{eq:33}) allows for performing
standard gaussian integrals and finally one obtains Eq.(\ref{eq:21})


\end{document}